\documentclass[prd,twocolumn,groupedaddress,showpacs]{revtex4}
\usepackage[dvips]{graphicx}
\usepackage[]{caption}
\usepackage{amsmath}
\usepackage{amssymb}
\begin{document}

\bibliographystyle{prsty}

\title{Non-Singular Perturbations in a Bouncing Brane Model}
\author{T.~J. Battefeld, S.~P. Patil, R. Brandenberger}
\affiliation{Physics Department, Brown University,
  Providence RI 02912 USA.}

\date{\today}

\begin{abstract}

The question of how perturbations evolve through a bounce in the Cyclic
and Ekpyrotic models of the Universe remains a topical one. Issues concerning singularities
at the background level and at the level of perturbation theory continue to be a matter of debate.
In this report we hope to demonstrate a non singular collision between the boundary
branes at the background level, and circumstances under which all perturbation variables
remain bounded through the collision.

As expected, we find most collisions to be singular
even in the full 5-D formalism, where first order perturbation theory breaks down for at least one
perturbation variable. Only in the case that the boundary branes
approach each other with constant velocity shortly before the bounce,
can a consistent, non singular solution be found. It is then possible
to follow the perturbations explicitly
until the actual collision. In this case, we find that if a scale
invariant spectrum developed on the hidden brane, it will get
transferred to the visible brane during the bounce.
\end{abstract}
\maketitle

\section{Introduction}

It has recently been proposed that the "Ekpyrotic" scenario of
the Universe \cite{KOST1,KOST2} or its revised version, the Cyclic
scenario \cite{cyclic}, could provide an alternative to inflationary
cosmology in providing an explanation for the observed spectrum
of fluctuations which is nearly scale-invariant and adiabatic
\cite{WMAP}. The effective four space-time dimensional picture of
the Ekpyrotic scenario corresponds to an initially contracting
Universe which undergoes a bounce and re-emerges in an expanding
phase. In the four-dimensional picture, the Ekpyrotic scenario
is closely related to the Pre-Big-Bang scenario \cite{PBB}
introduced several years earlier
\footnote{There are, however differences both in terms of the particle
physics motivation for the scenario - in the Ekpyrotic scenario the string
coupling goes to zero at the bounce, whereas in the Pre-Big-Bang scenario the
bounce occurs in the regime of strong string coupling - and also
in the details of the dynamics - the rate of contraction is different.}.
The Ekpyrotic scenario, however, is intrinsically five space-time
dimensional. Our Universe is a boundary brane in a five-dimensional
bulk. The phase of contraction (in the effective four-dimensional
picture) corresponds to the period in which a second brane (a bulk
brane in initial scenario \cite{KOST1}, the second boundary
brane in the later proposals \cite{KOST2,cyclic}) approaches our
brane. The bounce is the time of collision, and the phase of
expansion corresponds to the period when the distance between the
branes is once again increasing \footnote{See
\cite{pyro1,pyro2,rasanen2} for some criticism of the basic
scenario.}. The bounce was considered to be non-singular from
the five-dimensional perspective in the initial Ekpyrotic scenario,
but singular in the subsequent versions.

It has been shown \cite{KOST3}
that in the contracting phase of the Ekpyrotic scenario
quantum vacuum fluctuations generate a scale-invariant spectrum of
adiabatic cosmological fluctuations (more precisely, a scale-invariant
spectrum of perturbations of the Bardeen variable $\Phi$, the
fluctuation of the metric in longitudinal gauge - see \cite{MFB}
for a comprehensive review of the theory of cosmological perturbations,
and \cite{RHBrev} for a recent condensed version), and
that this scale-invariant spectrum would survive the bounce
(see \cite{PMPS} for a general criticism of the matching conditions
of \cite{KOST3} at the bounce). However, if one
uses standard matching conditions \cite{Hwang,Deruelle} on a constant
field hyper-surface to match the fluctuations in the contracting and
expanding phases, then one finds \cite{Finelli1} that the scale-invariant
mode of the contracting phase does not couple to the dominant mode of
the expanding phase, and that the dominant mode in the expanding phase
will inherit a phenomenologically completely unacceptable $n = 3$
spectrum (see also \cite{hwang2,tsujikawa}). Alternatively, one finds
that even in the contracting phase, the fluctuation variable $\zeta$,
which is the variable in terms of which the fluctuations are well-behaved
on super-Hubble scales in inflationary cosmology, has a blue $n = 3$
spectrum \cite{Lyth1}.

The question of how perturbations evolve through the bounce in
Ekpyrotic and Cyclic cosmologies (in terms of the bounce there
is no difference between the second Ekpyrotic scenario of \cite{KOST2}
and the cyclic model) is thus of central importance. In \cite{Durrer1}
it has been stressed that the spectrum of fluctuations obtained using
a matching surface analysis depends crucially of the exact choice of
the matching surface, and in \cite{Lyth2} it was pointed out that
in any coordinate system, one of the pair of canonical variables for
the fluctuations diverges, rendering the applicability of
any perturbative analysis questionable.

There have been recent attempts to - within an effective four-dimensional
picture - regularize the bounce and follow the fluctuations through
the bounce. One attempt \cite{Tsujikawa2} was to regularize the bounce
using higher derivative terms in the gravitational action (similar to
what was done for the Pre-Big-Bang scenario in \cite{Cartier}), with
results supporting a final $n = 3$ spectrum for the cosmological
fluctuations (see, however, the caveats discussed in \cite{Durrer2}).
In contrast, in the work of \cite{PP,Peter} it was found
that when modeling a nonsingular bounce using a spatially closed
Universe with matter obeying the usual energy conditions, the final
spectrum can depend crucially on the details of the bounce.

Therefore, it seems that only a fully five-dimensional analysis of
fluctuations can give the complete answer as to the final spectrum
of perturbations in the Ekpyrotic scenario. However, once again
the singularity of the bounce provides problems. A particular
matching condition, based on the assumption that the bounce in
the $t-y$ plane ($y$ being the coordinate perpendicular to the
branes) is a compactified Milne space, was recently studied in
\cite{Tolley2} (based on earlier work in \cite{Tolley1}
- see also \cite{Costa}),
yielding a scale-invariant
spectrum. On the other hand, the singularity of the bounce could
be more complicated than a simple Milne space (it could include
whisker regions), and in this case the influence of the whisker
regions on the post-bounce observer region would
destroy the scale-invariance of the spectrum
after the bounce (see e.g. \cite{Craps}).

Thus, in this work we will turn to a study of the evolution of
fluctuations in the five-dimensional model assuming that the
bounce is nonsingular at the level of the background cosmology.
We will thus assume that the collision is a
nonsingular event in 5D and only appears singular in an effective 4D
description. Note, however, that we will take the radius of the
extra spatial dimension to vanish at the time of the bounce, in
the spirit of the scenarios of \cite{KOST2,cyclic}, but unlike what
occurs in the original scenario of \cite{KOST1}, or in models in which
the bounce is smoothed out such that the branes never hit.
Then one can derive,
using homogeneity and isotropy in the three spatial dimensions, the
general form of the metric near the
bounce \cite{rasanen1}. Under the same assumptions, the field content
of the bulk can also be computed \cite{rasanen1,rasanen2}.

Given this background, we will perturb the metric and the bulk
matter fields. We will then try to
find a consistent, nontrivial solution to the perturbed equations of
motion, such that
\begin{enumerate}
\item The bounce remains nonsingular, that is the Riemann tensor stays
bounded.
\item Perturbation theory remains valid, i.e. all perturbed variables stay
bounded near the bounce.
\end{enumerate}
In the majority of cases, it will turn out that there is no such
solution for the cosmological fluctuations, and thus the
requirement that the perturbation variables remain bounded is
untenable. Therefore, the usual techniques
of evolving perturbations through the bounce are not applicable.
In fact, as a consequence of the divergence of the fluctuations, the
bounce itself will be a highly inhomogeneous and singular event
even when viewed from the 5D angle, possibly looking like a gas of
black holes (see e.g. \cite{BF,Veneziano} for recent ideas along
these lines).

The only consistent nonsingular background metric turns out to be
realized if the two branes approach each other with constant
velocity near the bounce. In that case we
will be able to explicitly compute how the
perturbations evolve during the bounce, so that there are no ambiguities
regarding matching perturbations before and after the bounce. We find
in this case that an initial bulk scale-invariant spectrum transfers
to a scale-invariant spectrum on the brane, thus supporting the
conclusions of \cite{Tolley2}.

The outline of this paper is as follows: After setting up a specific
realization of a bouncing Universe and
reviewing the background solution in Section \ref{sectionsetup},
we will perturb the metric and matter fields and identify the
scalar degrees of freedom and the resulting equations of motion
(EOM) (Section \ref{sectionperturb}). Thereafter
we search for a consistent solution, by first identifying constraints and
then working with the equations of motion directly
(Section \ref{sectionsearch}). We conclude
 with a discussion of the implications of our results for the Ekpyrotic
and Cyclic models of the Universe (Section
 \ref{sectionconsequences}).

If not stated otherwise the following notational conventions apply:
capital Latin indices run from 0 to 4,
small ones from 1 to 3 (spatial coordinates), and Greek indices from 0 to
3 (time and spatial). Repeated indices are to be summed over.
A dot denotes a partial derivative with respect to the bulk time $t$, a
prime indicates the partial derivative with respect to
the extra dimension $y$, and $\bigtriangleup$ is the 3-d flat space
Laplacian. We scale the time $t$ such that $t = 0$ is the time of the
collision of the branes.

\section{The setup\label{sectionsetup}}

To realize the Cyclic/Ekpyrotic scenario, we will work with the action
of heterotic M-theory \cite{HW}
\begin{eqnarray}
S=S_{het}+S_{BI}+S_{matter}\,,
\end{eqnarray}
where $S_{BI}$ describes the brane interaction that vanishes
(vanishing at least $\sim t$ near the bounce),
 $S_{matter}$ the matter content on the two
boundary branes at $y=y_1\equiv 0$ and $y=y_2$ (taken to be
empty before the bounce, and containing an
ideal fluid or scalar field matter after the bounce) and
\begin{eqnarray}
S_{het}&=&\frac{M_5^3}{2}\int _{\cal M}d^5x\sqrt{-g}
\Big(R-\frac{1}{2}\partial _A\Xi\partial ^A\Xi \label{action}\\
\nonumber &&-\frac{3}{2}
\frac{1}{5!}e^{2\Xi}{\cal F}_{ABCDE}{\cal F}^{ABCDE} \Big)
\\ \nonumber &&-\sum _{i=1}^2 3\alpha _i M^3_5
\int _{{\cal M}_4^{(i)}}d^4\xi _{(i)}
 \Big(\sqrt{-h_{(i)}}e^{-\Xi}\\
&& \nonumber -\frac{1}{4!}\epsilon ^{\mu\nu\kappa\lambda}{\cal A}_{ABCD}
\partial _\mu X_{(i)}^A
\partial _\nu X_{(i)}^B\partial _\kappa X_{(i)}^C
\partial _\lambda X_{(i)}^D\Big)
\end{eqnarray}
is a minimal version of the 5D heterotic M-theory action, with
$M_5$ the Planck mass in 5D,
$R$ the Ricci scalar in 5D,
$\Xi$ the dilaton, and $e^{\Xi}$ the volume of the Calabi-Yau threefold.
In addition, $A_{ABCD}$ is the four form gauge field with field strength
${\cal F}=d{\cal A}$,
$\alpha _i$ are the tensions of the visible and hidden branes for
$i=1,2$ respectively,
$g_{AB}$ is the metric on ${\cal M}_5={\cal M}_4\times S_1/Z_2$,
$h_{\mu\nu}^{(i)}$ is the induced metric
on the brane, and $X_{(i)}^A(\xi _{(i)}^\mu)$ are the brane embedding
coordinates.

The background metric $g_{AB}$ will have positive signature, that is
\begin{eqnarray}
ds^2\!&=&\!-n(t,y)^2dt^2\!+a(t,y)^2\delta_{ij}dx^i dx^j\!+b(t,y)^2dy^2\! .
\label{metrik}
\end{eqnarray}

\subsection{$Z_2$-symmetry}

As is already implicit from the form of the action $S_{het}$, we
are assuming a compactification of the five dimensional theory on the
orbifold $S_1/Z_2$. The visible and the hidden
brane are located at the orbifold fixed points at $y=y_1$ and $y=y_2$,
respectively. It should be noted that working on an
orbifold yields some additional constraints. For example, the metric has
to be $Z_2$-even and any infinitesimal
coordinate transformation $\xi ^A$ must be such that
$\xi ^y(y=y_i)=0$ in order to avoid delta-function singularities in the
metric (see \cite{brandenberger1} for a more detailed discussion).

\subsection{The Einstein equations}

{F}rom the metric (\ref{metrik}) it is straightforward to compute the
unperturbed Einstein equations,
which have to be satisfied by the background solution for the metric and
the fields $\Xi$ and ${\cal F}$ (see \cite{rasanen1}).

\subsection{Matter equations of motion}

Varying $S_{het}$ from (\ref{action}) with respect to the fields
$\Xi$ and ${\cal F}$, respectively, yields the EOM
\begin{eqnarray}
\nonumber 0&=&\Box \Xi
-\frac{3}{5!}e^{2\Xi}{\cal F}_{ABCDE}{\cal F}^{ABCDE}\\
&&+\sum _{i=1}^2\delta(y-y_i)b^{-1}6\alpha _i
e^{-\Xi}\,,\label{eomphi}\\
\nonumber 0&=&D_M\left(e^{2\Xi}{\cal F}^{MABCD}\right)\\
&&+\delta _0^{[A}\delta _1^{B}\delta _2^{C}\delta _3^{D]}\sum _{i=1}^2
\delta(y-y_i)(-g)^{-1/2}2\alpha _i\,.\label{eomf}
\end{eqnarray}

\subsection{Background solution}

The background metric is given by Eq. (\ref{metrik}).
We write each matter field as the sum of a spatially homogeneous
background field and a field representing small deviations around it
(and whose spatial average vanishes), that is:
\begin{eqnarray}
\nonumber \Xi(t,\vec{x},y)&=&\,^{(0)}\!\Xi(t,y) +^{(1)}\!\Xi(t,\vec{x},y)\\
&\equiv& ^{(0)}\!\Xi +\delta\Xi\,,\label{defphi}\\
\nonumber {\cal F}_{ABCDE}(t,\vec{x},y)
&=&\epsilon _{ABCDE}\left(\,^{(0)}\!{\cal F}(t,y)
+\,^{(1)}\!{\cal F}(t,\vec{x},y)\right)\\
&\equiv&\epsilon _{ABCDE}\left(\,^{(0)}\!{\cal F}
+\delta{\cal F}\right)\,.\label{deff}
\end{eqnarray}
Note that the Levi-Civita tensor implicitly contains the appropriate
metric factor so that it is a bona-fide tensor:
$\epsilon _{ABCDE}:=\sqrt{-g}\varepsilon _{ABCDE}$.

One can derive the general form of the unperturbed metric and the
unperturbed matter fields given above.
Assuming non-singularity, then near
the bounce the solutions can be written in terms of a power series
expansion in $t$ (see \cite{rasanen1,rasanen2}). One then takes the
following steps:
\begin{enumerate}
\item One demands that the bounce is non singular in 5D, i.e. that the
Riemann tensor $^{(0)}\!R_{ABCD}$,
 the Einstein tensor $^{(0)}\!G^A_{\,\,B}$ and the energy momentum tensor
$^{(0)}\!T^A_{\,\,B}$ remain bounded near the bounce.
\item One then solves the EOM of the matter fields as well as the Einstein
equations for the unperturbed metric for small $t$.
\item One enforces the matching conditions arising from delta functions,
i.e. one has to match the bulk solution
to the boundary values on the branes at $y=y_1$ and $y=y_2$.
\end{enumerate}
This finally yields the following possible forms for the metric:
\begin{eqnarray}
b(t,y)&=&Bt^k+B b_{k+1}(y)t^{k+1}+\mathcal{O}(t^{k+2})\,,
\label{metrikbbeforebounce}\\
\nonumber n(t,y)&=&1+\sum _{i=k}^{2k-3}\tilde{N}_iy t^i\\
\nonumber &&+\left(\frac{k(k-1)}{2}B^2y^2+\tilde{N}_{2k-2}y\right)
t^{2k-2}\\
&&+n_{2k-1}(y)t^{2k-1} +
\mathcal{O}(t^{2k})\,,\\
\nonumber a(t,y)&=&1+\sum _{i=0}^{k-1}A_it^i
+\sum _{i=k}^{2k-2}(\tilde{A}_iBy+A_i)t^i\\
\nonumber &&+\left(\frac{B^2A_1}{2}y^2+\tilde{A}_{2k-1}y+
A_{2k-1}\right)t^{2k-1}\\
&&+\mathcal{O}(t^{2k}) \,,
 \label{metrikabeforebounce}
\end{eqnarray}
with $k\geq 3$ and
\begin{eqnarray}
n_{2k-1}^{\prime\prime}(y)-B^2k(k+1)b_{k+1}(y)&=&0\,,
\end{eqnarray}
 and
$A_i$, $\tilde{A}_i$, $B$ and $\tilde{N}_i$
are constants. The $A_i$ may be zero and
$\tilde{N}_k=\tilde{A}_k=-\alpha _1 e^{-\Xi _0}/2$.

The other allowed metric with $k=1$ is \footnote{Note that this metric
does not correspond to a Milne like bounce,
since $n_1$ and $a_1$ depend on $y$.}
\begin{eqnarray}
b(t,y)&=&Bt+Bb_2(y)t^2+\mathcal{O}(t^3)\,,\label{metrikafterbounce1}\\
n(t,y)&=&1+n_1(y)t+\mathcal{O}(t^2)\,,\\
\nonumber a(t,y)&=&1+\left(\tilde {A}_1\sinh(By)
+A_1\cosh(By)\right)t+\mathcal{O}(t^2)\\
&\equiv&1+a_1(y)t+\mathcal{O}(t^2)\,,\label{metrikafterbounce3}
\end{eqnarray}
where $n_1$ and $b_2$ are related via
\begin{eqnarray}
 b_2(y)-\frac{n_1(y)}{2}-\frac{n^{\prime\prime}_1(y)}{2B^2}=0 \label{eomb2}
\end{eqnarray}
and the boundary conditions are
$n_1^{\prime}(y_i)=a_1^{\prime}(y_i)=-B\alpha _1 e^{-\Xi _0}/2$.
A mistake in \cite{rasanen1} regarding the case $k=1$ was
corrected in \cite{rasanen3}.
Before the bounce $k=1$ and $k\geq 3$ are possible, after the bounce only
the $k=1$ case is allowed. $B$ should have a different value before
and after the bounce.

Note that demanding in addition to the above boundedness requirements
a ``no flow condition'' (no energy is allowed to flow off the brane), then
the single allowed metric in the case $k=1$ after the bounce
(\ref{metrikafterbounce1})-(\ref{metrikafterbounce3})
is also ruled out \cite{rasanen1}. However, since in the Ekpyrotic
and Cyclic scenarios one assumes that energy can flow from the bulk to
the brane, it is not reasonable to demand this additional condition.

For the field $^{(0)}\!\Xi$  one gets by using its equation of motion
and the junction conditions
\begin{eqnarray}
\nonumber ^{(0)}\!\Xi(t,y) &=& \sum _{i=0}^{k-1}\Xi _it^i +
\left(3\alpha _1e^{-\Xi _0}By+\Xi _k\right)t^k\\
&&+\mathcal{O}(t^{k+1})
\end{eqnarray}
for $k\geq 3$ and
\begin{eqnarray}
\nonumber ^{(0)}\!\Xi(t,y) &=& \Xi _0+\left(\lambda \cosh(By) +
\tilde{\lambda}\sinh(By)\right)t\\
&&+\mathcal{O}(t^{2})
\end{eqnarray}
for $k=1$, where $\lambda ,\, \tilde{\lambda}$ and $\Xi _i$ are constants.
The field $^{(0)}\!{\cal F}$ is then given by
\begin{eqnarray}
^{(0)}\!{\cal F}(y,t) = \alpha _1e^{-2\,^{(0)}\!\Xi(y,t)}\,.
\end{eqnarray}

\section{Scalar Perturbations\label{sectionperturb}}

We will now perturb the metric focusing only on scalar metric perturbations.
The most general scalar perturbation in ``generalized'' longitudinal gauge
can be characterized by four scalar functions, $\Phi ,\Psi ,W$ and $\Gamma$
\cite{brandenberger1}. These functions can be viewed as a basis of gauge
invariant variables. In this gauge, the metric is given by
\begin{eqnarray}
\nonumber ds^2&=&-n(t,y)^2(1-2\Phi (t,\vec{x},y))\,dt^2\\
\nonumber &&+a(t,y)^2(1-2\Psi(t,\vec{x},y))\delta_{ij}dx^i dx^j\\
\nonumber &&+b(t,y)^2(1+2\Gamma(t,\vec{x},y))\,dy^2\\
&&-2n(t,y)^2W(t,\vec{x},y) \,dt\,dy \,,\label{perturbedmetrik}
\end{eqnarray}
where the signs in front of the perturbations are a mere convention.

Two more scalar degrees of freedom (matter degrees of freedom) were
already introduced in (\ref{defphi}) and (\ref{deff}). They are
$\delta \Xi(t,\vec{x},y)$ and $\delta{\cal F}(t,\vec{x},y)$.

\subsection{The brane embedding and longitudinal gauge}

It should be noted that we do not perturb the embedding of the boundary
branes - they are still localized at $y=y_1\equiv 0$ and $y=y_2$.
We are entitled to do this without loss of generality, since one can
induce local fluctuations in the radius by allowing $g_{55}$ to fluctuate,
rather than by making use of fluctuations in the coordinate interval.
However there are further compelling reasons for this
approach: perturbing the embedding would render the collision into a series
of local events in time, thus complicating the situation a lot.
This approach of not considering perturbations
in the position of the branes is common in the literature
(see e.g. \cite{Tolley2}). However, one should note
that due to the possible divergence of the metric at the
bounce, in general a large coordinate transformation would be necessary to
transform small perturbations in the coordinate position of the brane away.
This could induce further changes in the spectrum of perturbations
(keep in mind, that the gauge invariant variables introduced in
(\ref{perturbedmetrik}) are only gauge invariant under
infinitesimal coordinate transformations). Since we are only interested in
the physics of how a given perturbation evolves through the global bounce
(and not in exploring new ways for how the spectrum could be generated
at an inhomogeneous bounce) our simplified setup is satisfactory.

Note also that the dynamics of the collision is fully included in the
metric - the brane coordinates are time-independent
but the physical distance between the branes does go to zero when
$t$ approaches zero.

One may wonder if keeping the branes at fixed positions is consistent
with working in the generalized longitudinal gauge. First note that the
branes sit at the orbifold fixed points - demanding that the metric
may not include delta function singularities requires an infinitesimal
coordinate transformation $\xi ^A$ to be continuous
everywhere, specifically $\xi _y(y=y_i)\equiv 0$ has to hold at the brane
positions \cite{brandenberger1}.
This ensures that the coordinate transformation necessary to go to the
longitudinal gauge does not perturb the brane position.

It is clear that this situation depends on our choice to work with branes
localized at orbifold fixed planes. As discussed e.g. in \cite{VDB},
in a more general brane world setting one needs to include an extra
degree of freedom for scalar metric fluctuations. In our setting,
going to longitudinal gauge and keeping the brane positions fixed does not
introduce any new restrictions on the perturbations considered.

\subsection{The perturbed Einstein equations}

To write down the Einstein equations we need both the perturbed
energy-momentum tensor $^{(1)}T^A_{\,\,B}$ and
the perturbed Einstein tensor $^{(1)}G^A_{\,\,B}$, computed up to first
order in the perturbation variables. The latter is straightforward to
compute from (\ref{perturbedmetrik}) and reads
in component form (no summation convention where repeated indices appear)
\begin{eqnarray}
^{(1)}G^{x_i}_{\,\,x_j}&=&\frac{1}{a^2}\partial _{x_i}\partial _{x_j}
\left[\Psi +\Phi -\Gamma\right]\,,\label{G^i_j;ineqj}
\end{eqnarray}

\begin{widetext}
\begin{eqnarray}
^{(1)}G^t_{\,\,t}&=&\Bigg[\frac{3}{b^2}\left(\frac{b^\prime}{b}\partial _y
-4\frac{a^\prime}{a}\partial _y-\partial ^2_y\right)
+\frac{3}{n^2}\left(\frac{\dot{b}}{b}\partial _t
+2\frac{\dot{a}}{a}\partial _t\right)-\frac{2}{a^2}\triangle\Bigg]\Psi
-\frac{6}{n^2}\left(\frac{\dot{a}\dot{b}}{ab}+\frac{\dot{a}^2}{a^2}\right)
\Phi \label{einsteintt}\label{G^t_t}\\
\nonumber &&\Bigg[\frac{3}{b^2}
\left(-2\frac{a^{\prime 2}}{a^2}-2\frac{a^{\prime\prime}}{a}
+2\frac{a^{\prime}b^\prime}{ab}
-\frac{a^{\prime}}{a}\partial _y\right)-\frac{3\dot{a}}{n^2a}\partial _t
+\frac{1}{a^2}\triangle\Bigg]\Gamma +\frac{3}{b^2}\left(-\frac{\dot{a}^\prime}{a}
-\frac{n^\prime\dot{a}}{na}-2\frac{\dot{a}a^\prime}{a^2}
+\frac{\dot{a}b^\prime}{ab}-\frac{\dot{a}}{a}\partial _y\right)W \,,\\
^{(1)}G^{x_i}_{\,\,t}&=&\partial _{x_i} \Bigg[\frac{2}{a^2}\partial _t\Psi
-\frac{1}{a^2}\left(\frac{\dot{b}}{b}+2\frac{\dot{a}}{a}\right)\Phi
-\frac{1}{a^2}\left(\frac{\dot{b}}{b}-\frac{\dot{a}}{a}+\partial _t\right)\Gamma
-\frac{1}{2a^2b^2}\left(3nn^\prime +n^2\frac{a^\prime}{a}-n^2\frac{b^\prime}{b}
+n^2\partial _y\right)W\Bigg] \,, \label{einsteinxit}\\
\nonumber ^{(1)}G^y_{\,\,t}&=&\frac{3}{b^2}
\left(\frac{\dot{a}}{a}\partial _y-\frac{\dot{b}}{b}\partial _y
+\frac{a^\prime}{a}\partial _t-\frac{n^\prime}{n}
\partial _t+\partial _y\partial _t\right)\Psi
-\frac{3\dot{a}}{b^2a}\partial _y\Phi
+\frac{6}{b^2}\left(\frac{\dot{a}^\prime}{a}
-\frac{a^\prime \dot{b}}{ab}-\frac{n^\prime\dot{a}}{na}
+\frac{a^\prime}{2a}\partial _t\right)\Gamma\\
&&+\left[\frac{3}{b^2}\left(-\frac{\dot{n}\dot{a}}{na}
+\frac{\ddot{a}}{a}-\frac{\dot{a}\dot{b}}{ab}\right)
+\frac{n^2}{2b^2a^2}\triangle\right]W \,,\label{einsteinyt}\\
 ^{(1)}G^{x_i}_{\,\,x_i}&=&\frac{1}{a^2}
 \left[\partial _{x_i}\partial _{x_i}-\triangle\right]\left(
-\Gamma +\Psi +\Phi \right)
+\left[\frac{2}{n^2}\left(-\frac{\dot{n}}{n}+\frac{\dot{b}}{b}
+3\frac{\dot{a}}{a}+\partial _t\right)
\partial _t-\frac{2}{b^2}\left(3\frac{a^\prime}{a}
-\frac{b^\prime}{b}+\frac{n^\prime}{n}+\partial _y\right)
\partial _y\right]\Psi \label{einsteinxixi}\\
\nonumber &&\!\!\!\!\!\!-\Bigg[\frac{2}{n^2}\left(\frac{\ddot{b}}{b}
-2\frac{\dot{n}\dot{a}}{na}+2\frac{\dot{a}\dot{b}}{ab}
+\frac{\dot{a}^2}{a^2}-\frac{\dot{n}\dot{b}}{nb}+2\frac{\ddot{a}}{a}
+\frac{\dot{b}}{2b}\partial _t
+\frac{\dot{a}}{a}\partial _t\right)
+\frac{1}{b^2}\left(-\frac{b^\prime}{b}
+2\frac{n^\prime}{n}+2\frac{a^\prime}{a}+
\partial _y\right)
\partial _y\Bigg]\Phi\\
\nonumber &&\!\!\!\!\!\!-\Bigg[\frac{2}{b^2}\left(2\frac{a^{\prime\prime}}{a}
-2\frac{a^\prime b^\prime}{ab}
-\frac{n^\prime b^\prime}{nb}+\frac{a^{\prime 2}}{a^2}
+\frac{n^{\prime\prime}}{n}+2\frac{n^{\prime}a^\prime}{na}
+\frac{a^{\prime}}{a}\partial _y+\frac{n^\prime}{2n}\partial _y\right)
+\frac{1}{n^2}\left(2\frac{\dot{a}}{a}
-\frac{\dot{n}}{n}+2\frac{\dot{b}}{b}+\partial _t
\right)
\partial _t\Bigg]\Gamma\\
\nonumber &&\!\!\!\!\!\!-\frac{2}{b^2}\Bigg[\frac{a^\prime\dot{a}}{a^2}
+\frac{\dot{n}^\prime}{n}+2\frac{\dot{a}^\prime}{a}
+\frac{\dot{n}a^\prime}{na}-\frac{\dot{n}b^\prime}{2nb}
+\frac{\dot{a}n^\prime}{an}-\frac{\dot{b}n^\prime}{2bn}
-\frac{b^\prime \dot{a}}{ba}-\frac{\dot{b}a^\prime}{ba}
+\left(\frac{a^\prime}{a}-\frac{b^\prime}{2b}+\frac{n^\prime}{n}
\right)\partial _t
+\left(\frac{\dot{a}}{a}+\frac{\dot{n}}{2n}
+\frac{\partial _t}{2}\right)\partial _y
\Bigg]W \,, \\
 ^{(1)}G^{y}_{\,\,x_i}&=&\partial _{x_i}\Bigg[\frac{2}{b^2}\partial _y\Psi
+\frac{1}{b^2}\left(-\frac{a^{\prime}}{a}+\frac{n^\prime}{n}
+\partial _y\right)\Phi
+\frac{1}{b^2}\left(2\frac{a^\prime}{a}+\frac{n^\prime}{n}\right)\Gamma
+\frac{1}{2b^2}\left(\frac{\dot{b}}{b}+\frac{\dot{a}}{a}
+\frac{\dot{n}}{n}+\partial _t\right)W\Bigg]\label{einsteinyxi}
\,,\\
^{(1)}G^{y}_{\,\,y}&=&\left[-\frac{2}{a^2}\triangle +\frac{3}{n^2}\left(\partial _t
-\frac{\dot{n}}{n}+4\frac{\dot{a}}{a}\right)\partial _t
+\frac{3}{b^2}\left(-2\frac{a^\prime}{a}-\frac{n^\prime}{n}\right)
\partial _y\right]\Psi
+\frac{6}{b^2}\left(-\frac{a^{\prime 2}}{a^2}
-\frac{a^\prime n^\prime}{an}\right)\Gamma
\\
\nonumber && +\left[-\frac{1}{a^2}\triangle
-\frac{6}{n^2}\left(\frac{\dot{a}^2}{a^2}-\frac{\dot{n}\dot{a}}{na}
+\frac{\ddot{a}}{a}+\frac{\dot{a}}{2a}\partial _t\right)
-\frac{3}{b^2}\frac{a^\prime}{a}\partial _y\right]\Phi
-\frac{3}{b^2}\left(\frac{\dot{n}a^\prime}{na}
+2\frac{a^\prime\dot{a}}{a^2}-\frac{a^\prime\dot{b}}{ab}
+\frac{\dot{a}^\prime}{a}
+\frac{a^\prime}{a}\partial _t\right)W \,. \label{einsteinyy}
\end{eqnarray}
\end{widetext}
The right hand side of the Einstein equations is given by the perturbed
energy-momentum tensor, which can be derived in
a straightforward way by varying the action with respect to the
metric and keeping only first order contributions. One finds that
\begin{widetext}
\begin{eqnarray}
\frac{^{(1)}\!T^{A}_{\,\,B}}{M_5^3}
&=&-\frac{\delta ^A_B}{2}\partial _C\,^{(0)}\!\Xi
\partial ^C\delta\Xi +\frac{1}{2}\partial ^A\delta\Xi
\partial _B\,^{(0)}\!\Xi +\frac{1}{2}
\partial ^A\,^{(0)}\!\Xi\partial _B\delta\Xi
+\delta ^A_B\frac{3}{2}\frac{1}{5!}e^{2\,^{(0)}
\!\Xi}\,^{(0)}\!{\cal F}_{CDEFG}\,^{(1)}\!{\cal F}^{CDEFG}
\label{perturbedenergymomentum}\\
\nonumber &&+\delta ^A_B \frac{3}{2}\frac{1}{5!}e^{2\,^{(0)}\!\Xi}
\delta\Xi \,^{(0)}\!{\cal F}_{CDEFG}\,^{(0)}\!{\cal F}^{CDEFG}
+\delta _{B4}\frac{1}{M_5^2}\sum _{i=1}^2\frac{\delta(y-y_i)}{b(y_i)}
\delta g_{B\beta}T^{A\beta}_{(i)}\\
\nonumber &&+\frac{1}{M_5^3}\sum _{i=1}^2\frac{\delta(y-y_i)}{b(y_i)}
\Bigg[\frac{\delta g_{44}}{2b^2}\left(
\delta ^A_B e^{-\,^{(0)}\!\Xi}3\alpha _i M_5^3-T^{A}_{(i)B}\right)
+\delta g_{B\beta}T_{(i)}^{A\beta}+\delta ^A_Be^{-\,^{(0)}\!\Xi}3
\alpha _iM_5^3\delta\Xi
+g_{B\beta}\delta T_{(i)}^{A\beta} \Bigg]\,,
\end{eqnarray}
\end{widetext}
where the last sum is only present if $A\neq 4$ and $B\neq 4$.

>From the $x_i$-$x_j$-component of the perturbed Einstein equations
it follows that the the $x_i$-$x_j$-component of $^{(1)}\!G^{A}_{\,\,B}$
vanishes, as can be seen from (\ref{perturbedenergymomentum}).
Thus from (\ref{G^i_j;ineqj}) we have
\begin{eqnarray}
\frac{1}{a^2}\partial _{x_i}\partial _{x_j}\left[\Psi +\Phi -\Gamma\right]=0\,.
\end{eqnarray}
By exploiting rotational invariance in the three spatial directions,
and requiring that the perturbations remain bounded everywhere
and have a vanishing spatial average one can deduce that
$\Gamma = \Psi + \Phi$ up to a function of $t$ and $y$ \footnote{One can
write the perturbed metric in a rotated coordinate frame. The
infinitesimal perturbation variables do not change, and hence we can
conclude that all second
order 4D spatial derivatives of the off-diagonal Einstein tensor vanish.}.
Thus,
\begin{eqnarray}
\Gamma(t,\vec{x},y)=\Phi(t,\vec{x},y)+\Psi(t,\vec{x},y)+f(t,y)\,,
\end{eqnarray}
where $f(y,t)$ is a small, so far undetermined function whose average over
the extra dimension has to vanish. Thus, for
any spatial Fourier mode except the zero mode we may write
\begin{eqnarray}
\Gamma(t,\vec{q},y)=\Phi(t,\vec{q},y)+\Psi(t,\vec{q},y) \, ,
\end{eqnarray}
but the zero mode still contains the function $f(t,y)$. Since we are doing
linear perturbation theory, all Fourier modes decouple and
thus $f(t,y)$ is of no importance for the spectrum of metric fluctuations.
However, it may be important for the evolution on the brane,
since $f(t,y=y_i)$ may look like a small, time-varying cosmological
constant on the brane. This subtlety is often glanced over in the
literature \cite{brandenberger1}.

\subsection{The perturbed matter equations of motion}
To get the perturbed equations of motion we have to plug in the
perturbed metric from (\ref{perturbedmetrik}) into
(\ref{eomphi}) and (\ref{eomf}), use the definitions (\ref{defphi})
and (\ref{deff}), and expand the resulting equations
up to first order in the perturbations. For the smooth component of the
fields we obtain
\begin{eqnarray}
\nonumber 0&=&^{(0)}\!\Box \delta\Xi +\,^{(1)}\!\Box^{(0)}\Xi \\
&&+4\left(\delta \Xi\, ^{(0)}\!{\cal F}^2 +\, ^{(0)}\!{\cal F}
\delta {\cal F}\right)e^{2\,^{(0)}\!\Xi}
\,,\label{eomphi1}\\
0&=&\partial _\mu\left(\left[\,^{(0)}\!{\cal F}2\delta\Xi+\delta{\cal F}
\right]e^{2\,^{(0)}\!\Xi}\right)\,.\label{eomf1}
\end{eqnarray}
We will return to the delta function part of (\ref{eomphi1}) in
section \ref{sectionconstraints}, where we
will extract constraints from it.
$^{(0)}\!\Box$ and $^{(1)}\!\Box$ are straightforward
to compute using the definition
\begin{eqnarray}
\nonumber \Box&\equiv&g^{AB}\partial _A\partial _B -g^{BC}\Gamma ^A_{BC}\partial _A\\
&=:&\,^{(0)}\!\Box+\,^{(1)}\!\Box +{\cal O}(\mbox{perturbations}^2)
\end{eqnarray}
and the perturbed metric (\ref{perturbedmetrik}). One gets
\begin{eqnarray}
\nonumber ^{(0)}\!\Box&=&-\frac{1}{n^2}\left[\partial _t^2
+\left(-\frac{\dot{n}}{n}+\frac{\dot{b}}{b}+3\frac{\dot{a}}{a}\right)
\partial _t\right]+\frac{1}{a^2}\triangle \\
&&+\frac{1}{b^2}\left[\partial _y^2+\left(\frac{n^\prime}{n}
-\frac{b^\prime}{b}+3\frac{a^\prime}{a}\right)\partial _y\right]\,,\\
\nonumber ^{(1)}\!\Box&=&2n^2\Phi\partial _t^2-2a^2\Psi\triangle
+2b^2\Gamma\partial _y^2-2n^2W\partial _y\partial _t\\
\nonumber&& +\Bigg\{\frac{1}{n^2}\left[\left(-2\frac{\dot{n}}{n}
+6\frac{\dot{a}}{a}+2\frac{\dot{b}}{b}\right)\Phi
+\dot{\Phi}-3\dot{\Psi}+\dot{\Gamma}\right]\\
\nonumber &&+\frac{1}{b^2}\left[\left(\frac{n^\prime}{n}
-\frac{b^\prime}{b}\right)W
+W^\prime\right]\Bigg\}\partial _t \\
\nonumber &&+\frac{1}{b^2}\Bigg[\left(2\frac{n^\prime}{n}
-2\frac{b^\prime}{b}+6\frac{a^\prime}{a}\right)\Gamma +
\Gamma ^\prime +\Phi ^\prime +3\Psi ^\prime\\
\nonumber && +\left(-\frac{\dot{b}}{b}+\frac{\dot{n}}{n}
+3\frac{\dot{a}}{a}\right)W
+\dot{W}\Bigg]\partial _y\\
&&+\frac{1}{a^2}\sum_{i=1}^3\left(\Phi _{,i}+\Psi _{,i}
-\Gamma _{,i}\right)\partial _{x_i}\,,
\end{eqnarray}
where $\triangle$ is the usual spatial Laplacian.

Note that (\ref{eomf1}) is in fact a algebraic equation relating
$\delta{\cal F}$ to $\delta \Xi$, since a simple integrations yields
\begin{eqnarray}
\delta{\cal F}=-2\,^{(0)}\!{\cal F}\delta{\Xi}\label{kickoutF}\,,
\end{eqnarray}
where the integration constant was set to zero since
this is the conclusion one would arrive at from solving for the gauge
field in terms of the dilaton through the equations of motion \cite{rasanen1}.

\section{In search of a consistent solution \label{sectionsearch}}

Now that we have introduced our notation and equipped ourselves with all
the relevant equations of motion we will try
to find a self consistent, nontrivial solution. Let us first make a
general ansatz for all perturbed
quantities and try to exploit all constraints before plugging the ansatz
into the respective equations.

There are three main sources of constraints:
\begin{enumerate}
\item In order to be consistent with our background solution, we must
demand that the perturbed Riemann tensor
$\,^{(1)}\!R_{ABCD}$ remains bounded near the bounce.
\item The perturbed energy-momentum tensor $\,^{(1)}T^A_{\,\,B}$ should
remain bounded.
\item The delta function parts of the EOM have to be matched with
possible divergent parts
originating from derivatives with respect to the extra dimension.
\end{enumerate}

Assuming that the fields describing the fluctuations do not diverge
at the bounce $t = 0$, we write down a general power law expansion
for them (following the philosophy that was used for the background
fields). We will then check as to whether nontrivial solutions are
possible. If not, we can then conclude that the perturbations
diverge at the bounce. Thus, we make the ansatz
\begin{eqnarray}
\Phi(t,\vec{x},y)&=&\varphi _0(\vec{x},y)+\varphi _s(\vec{x},y)t^s
+\mathcal{O}(t^{s+1})\,,\label{ansatzfields1}\\
\Psi(t,\vec{x},y)&=&\psi _0(\vec{x},y)+\psi _q((\vec{x},y))t^q
+\mathcal{O}(t^{q+1})\,,\\
W(t,\vec{x},y)&=&w_0(\vec{x},y)+w_l(\vec{x},y)t^l
+\mathcal{O}(t^{l+1})\,,\\
\delta\Xi(t,\vec{x},y)&=&\xi _0(\vec{x},y)+\xi _m(\vec{x},y) t^m
+\mathcal{O}(t^{m+1})\,,\label{ansatzfields6}
\end{eqnarray}
where $s,q,l$ and $m$ are the positive exponents of the leading
order contributions in $t$.

If we want the bulk perturbations to be able to seed nontrivial brane
fluctuations during the bounce, we require the additional condition
$\varphi _0(\vec{x},y)\neq 0$ and/or $\psi _0(\vec{x},y)\neq 0$ if $k\geq 3$.
The possible non-trivial case that the perturbations vanish but
their time derivative does not
will be ruled out shortly, since it turns out that
$q\geq k$ and $s\geq k $, with $k\geq 3$.

Note that if this ansatz is consistent and nontrivial near the bounce,
any perturbation variable evaluated on the visible brane at the bounce is
trivially determined by its bulk value shortly before the bounce.

This is consistent with the physical picture of the Ekpyrotic/Cyclic scenarios:
Perturbations are generated as quantum vacuum fluctuations of the bulk
scalar modes well before the collision of the branes (pictorially,
they can be represented as fluctuations in the distance between the
branes). They evolve as the two branes approach each other until shortly
before the bounce. The key question is how  the initial spectrum
of fluctuations in the ``contracting'' phase (the phase during which the
branes are approaching each other) is transferred to the spectrum of
perturbations on the brane immediately after the bounce,
assuming that there are no pre-existing fluctuations on the brane.
In the Ekpyrotic/Cyclic scenarios,
the background evolution is chosen such that this bulk spectrum is
scale-invariant \footnote{In these scenarios, the mode of $\Phi$ with
a scale-invariant spectrum is growing in the contracting phase so that
at the time of the bounce the magnitude of the fluctuations is much
larger than the magnitude of any pre-existing quantum vacuum fluctuations
on the brane.}. Our analysis, however, is independent of the origin and
shape of the initial bulk spectrum.

\subsection{Before the bounce}
There are two possibilities $k=1$ and $k\geq 3$. We will first work out
some general constraints before we
study these two cases separately. In addition we will only focus on the
spatially inhomogeneous modes
of the perturbations, so all results should be taken modulo a possible
non-vanishing zero Fourier mode.

\subsubsection{Boundedness of the Riemann tensor}
Demanding that potentially divergent terms in $\,^{(1)}\!R_{ABCD}$ get
canceled, we obtain the following constraints on the exponents:
\begin{eqnarray}
l&\geq& k+1\,,\\
s&\geq&k\,,\\
q&\geq&k\,.
\end{eqnarray}
Furthermore, the functions in our ansatz
(\ref{ansatzfields1})-(\ref{ansatzfields6}) are constrained to
\begin{eqnarray}
\varphi _0(\vec{x},y)&\equiv&\varphi _0(\vec{x})\,,\label{constraintrieman1}\\
\psi _0(\vec{x},y)&\equiv&\psi _0(\vec{x})\,,\label{constraintrieman2}\\
w_0(\vec{x},y)&\equiv&0\,.\label{constraintrieman3}
\end{eqnarray}
Deriving this is a straightforward exercise - one simply writes down all
components of the Riemann tensor to all orders in $t$ until
${\mathcal O}(t^{-1})$ and demands that they vanish.

To be specific, let us retrace the main steps leading to the constraints in
the case that $k\geq 3$: $\,^{(1)}\!R_{tx_3x_3y}$
yields to order ${\mathcal O}(t^{-1})$
\begin{eqnarray}
\psi _0^\prime +\frac{A_1}{A_0}w_0=0 \label{exampel1}\,.
\end{eqnarray}
On the other hand, $\,^{(1)}\!R_{tx_1x_1x_i}$ yields to order
${\mathcal O}(t^{-j})$,
\begin{eqnarray}
\frac{A\tilde{A}_k}{B^2}\partial _{x_i}w_0=0\,,
\end{eqnarray}
so that $w_0$ can not depend on $\vec{x}$. Therefore $w_0$ only couples to
the zero Fourier modes,
which we are not interested in. Thus, we may set $w_0\equiv 0$,
and from (\ref{exampel1}) we get consequently
$\psi _0(\vec{x},y)\equiv \psi _0(\vec{x})$. Stepping up in orders of
$t$ in $\,^{(1)}\!R_{tx_1x_1x_i}$ to ${\mathcal O}(t^{-j})$,
where $1\leq j \leq k$, we obtain
\begin{eqnarray}
\partial _{x_i}w_j=0\,,
\end{eqnarray}
so that $l\geq k$. Now, looking at $\,^{(1)}\!R_{tx_1tx_1}$ to order
${\mathcal O}(t^{-k})$ we get, using the previous results,
$\varphi _0(\vec{x},y)\equiv \varphi _0(\vec{x})$. Turning
to $\,^{(1)}\!R_{x_1x_2x_1x_2}$ we obtain to leading order
\begin{eqnarray}
2A_0\tilde{A}_k\psi _q^{\prime}t^q-2A_1\tilde{A}_kw_lt^l={\mathcal O}(t^k)\,,
\end{eqnarray}
so that $q\geq k$ has to hold. As a consequence, it follows from
$\,^{(1)}\!R_{tx_1tx_1}$
\begin{eqnarray}
\varphi _s ^{\prime}t^s={\mathcal O}(t^k)\,,
\end{eqnarray}
so that $s\geq k$ has to hold too. Finally, if we consider the
$y$-$x_i$-Einstein equation to order ${\mathcal O}(t^{-(k+1)})$, we see that
\begin{eqnarray}
\partial _{x_i}w_k={\mathcal O}(t^{k+1})
\end{eqnarray}
must be satisfied so that $l\geq k+1$ -- all in all we have derived the
claimed constraints in the case $k\geq 3$. No
other component of the Riemann tensor gives additional constraints.
The arguments for $k=1$ are similar.

\subsubsection{Boundedness of the energy momentum tensor}
Looking at
\begin{eqnarray}
T^y_{\,\,x_i}=\partial _{x_i}\left(\frac{1}{2b^2}
\delta\Xi \partial _y\,^{(0)}\!\Xi\right)
\end{eqnarray}
we see immediately that
\begin{eqnarray}
\xi _0(\vec{x},y)&\equiv& 0\,,\\
m&\geq&k
\end{eqnarray}
has to hold in order for the energy-momentum tensor to remain bounded.
The other components do not give any additional constraints.

\subsubsection{Matching to the branes\label{sectionconstraints}}
The next conditions we should look at originate from matching
$\delta$-function parts in the Einstein equations and the
EOM of the field $\delta\Xi$. Note that the
background solutions already satisfy similar constraints including only
unperturbed quantities.

Let us first look at the Einstein tensor: the only possible divergent terms
consist of second order derivatives
with respect to $y$. Thus, from (\ref{G^t_t})--(\ref{einsteinyy}) we see
that only the $t$-$t$ and $x_i$-$x_i$ components
of the Einstein tensor contain such terms. These parts are, specifically,
\begin{eqnarray}
^{(1)}\!G^t_{\,\,t}\big|_\delta&=&-\frac{3}{b^2}\Psi ^{\prime\prime}
-\frac{6}{b^2}\frac{a^{\prime\prime}}{a}\Gamma\,,
\label{Gdelta1}\\
^{(1)}\!G^{x_i}_{\,\,x_i}\big|_\delta&=&
-\frac{2}{b^2}\left[\Psi ^{\prime\prime}+\frac{\Phi ^{\prime\prime}}{2}
+\left(\frac{2a^{\prime\prime}}{a}
+\frac{n^{\prime\prime}}{n}\right)\Gamma\right]. \label{Gdelta2}
\end{eqnarray}
Note that the summation convention is not used in the above expression
for the $x_i$-$x_i$ component. Let us introduce some notation \cite{binetruy1}:
Splitting
$X^{\prime\prime}$ into a continuous part $\hat{X}^{\prime\prime}$
and a delta function part multiplied by the
jump in the first derivative $[X^{\prime}]$, that is, e.g.
\begin{eqnarray}
\Phi ^{\prime\prime}=\hat{\Phi}^{\prime\prime}+[\Phi ^{\prime}]\delta(y-y_i)\,.
\end{eqnarray}
Now, exploiting the $Z_2$-symmetry of the bulk we know the jump of the
perturbation variables. For example, in the case of the
brane at the origin we know
\begin{eqnarray}
\Phi ^{\prime}(t,\vec{x},y=0^+)&=&-\Phi ^{\prime}(t,\vec{x},y=0^-)\,,\\
\Psi ^{\prime}(t,\vec{x},y=0^+)&=&-\Psi ^{\prime}(t,\vec{x},y=0^-)\,,\\
\Gamma ^{\prime}(t,\vec{x},y=0^+)&=&-\Gamma ^{\prime}(t,\vec{x},y=0^-)\,,\\
W^{\prime}(t,\vec{x},y=0^+)&=&W^\prime(t,\vec{x},y=0^-)\,.
\end{eqnarray}
Thus, e.g. for $\Phi ^{\prime}$ at $y=0$ we have
\begin{eqnarray}
\nonumber [\Phi ^{\prime}]&\equiv&\Phi ^{\prime}(t,\vec{x},y=0^+)
-\Phi ^{\prime}(t,\vec{x},y=0^-)\\
&=&2\Phi ^{\prime}(t,\vec{x},y=0^+)\,.
\end{eqnarray}
Note that $a,b$ and $n$ behave like $\Phi$ at the brane positions
under the $Z_2$-symmetry.
With the perturbed energy-momentum tensor of the form
(\ref{perturbedenergymomentum}) with
(\ref{Gdelta1}) and (\ref{Gdelta2}) we get our first two constraints
\begin{eqnarray}
\nonumber\frac{3}{b}e^{-\,^{(0)}\!\Xi}\alpha _i(\Gamma
+\delta\Xi)+\mathcal{O}(t)&=&
 -\frac{3}{b^2}[\Psi ^{\prime}]-\frac{6}{b^2}\frac{[a^{\prime}]}{a}\Gamma
\,,\\
\nonumber\frac{3}{b}e^{-\,^{(0)}\!\Xi}\alpha _i(\Gamma
+\delta\Xi)+\mathcal{O}(t)
&=&-\frac{2}{b^2}[\Psi ^{\prime}]-\frac{1}{b^2}[\Phi ^{\prime}]\\
\nonumber &&-\frac{2}{b^2}\left(2\frac{[a^{\prime}]}{a}+
\frac{[n^{\prime}]}{n}\right)\Gamma\,,
\end{eqnarray}
which have to be satisfied on the branes at $y=0$ and $y=y_2$.
We have made use of the assumption that there is
no matter present on the branes before the collision. The physical picture
is, of course, that matter is created during the collision by the transfer
of kinetic energy of the hidden brane to the visible brane,
but for our purposes it will be satisfactory to simply
assume a sudden creation of matter at $t=0$, so that we have some
ideal fluid present on the visible brane for $t>0$ (naturally, this
assumption does effect the matching conditions for positive $t$).

A more convenient form of these two constraints is obtained by
taking the difference and the sum, yielding
\begin{eqnarray}
\mathcal{O}(t)&=&\frac{1}{b^2}\Bigg\{[\Phi ^\prime]-[\Psi ^\prime]
-2\left(\frac{[a^\prime]}{a}
-\frac{[n^\prime]}{n}\right)\Gamma \Bigg\}
\,,\label{constraint1}
\end{eqnarray}
and
\begin{eqnarray}
\frac{6}{b}e^{-\,^{(0)}\!\Xi}\alpha _i(\Gamma +\delta\Xi)+\mathcal{O}(t)&=&
-\frac{1}{b^2}\Bigg\{[\Phi ^\prime]+5[\Psi ^\prime]\\
\nonumber &&+2\left(5\frac{[a^{\prime}]}{a}+\frac{[n^{\prime}]}{n}\right)
\Gamma\Bigg\}
\,.\label{constraint2}
\end{eqnarray}
We obtain one more constraint from the delta function part of the
equation of motion for $\delta\Xi$ in (\ref{eomphi}),
expanded to first order in the perturbations.
One gets, e.g., for the brane at $y=0$
\begin{eqnarray}
\frac{\delta\Xi ^\prime}{b} -3\alpha _1 \left(\delta\Xi
+\frac{\Gamma}{2}\right)e^{-\,^{(0)}\!\Xi}&=&\mathcal{O}(t)
\,.\label{constraint3}
\end{eqnarray}

\subsubsection{The Einstein Equations for $k\geq 3$}
In case $q>3$, it follows from the junction conditions (\ref{constraint1})
and (\ref{constraint2}) that
\begin{eqnarray}
s&>&3\,,\\
\varphi _0(\vec{x})&=&-\psi _0 (\vec{x})\,.
\end{eqnarray}
In case when $q=3$, all we can deduce from the junction conditions alone is
\begin{eqnarray}
s&=&3\,,\\
\nonumber \psi _k^{\prime}(\vec{x},y=y_i)&\equiv&\varphi _s^{\prime}(\vec{x},y=y_i)\\
&\sim&(\varphi _0(\vec{x})+\psi _0(\vec{x}))\label{b1}\,.
\end{eqnarray}
Now, looking at the $x_i$-$t$ Einstein equation (\ref{einsteinxit}) to
leading order in $t$, we see that $w_{k+1}^{\prime}(\vec{x},y)\equiv 0$.
Using this knowledge in the $y$-$x_i$ equation (\ref{einsteinyxi}), we get
\begin{eqnarray}
w_{k+1}(\vec{x},y)\sim(\varphi _0(\vec{x})+\psi _0(\vec{x}))\label{b2}\,.
\end{eqnarray}
Working out the $y$-$t$ equation (\ref{einsteinyt}) to order
$\mathcal{O}(t^{-k})$ and evaluating
it at the brane we obtain, after using (\ref{b1}) and (\ref{b2}),
\begin{eqnarray}
\varphi _0(\vec{x})&=&-\psi _0 (\vec{x})\,.
\end{eqnarray}
Thus, in any case we can deduce $l>k+1$ as well as
\begin{eqnarray}
\varphi _0(\vec{x})&=&-\psi _0(\vec{x})
\end{eqnarray}
and
\begin{eqnarray}
\mbox{if}\,\,q=k\Rightarrow\,\,&\psi _q^{\prime}(\vec{x},y=y_i)&= 0\,,\\
\mbox{if}\,\,s=k\Rightarrow\,\,&\varphi _s^{\prime}(\vec{x},y=y_i)&= 0\,.
\end{eqnarray}
Going back to the junction condition for $\delta\Xi$, it follows that
\begin{eqnarray}
\mbox{if}\,\,m=k\Rightarrow\,\,\xi _m^{\prime}(\vec{x},y=y_i)= 0\,.
\end{eqnarray}
Equipped with this knowledge, it can be seen that the leading
order contribution to the $y$-$y$ equation (\ref{einsteinyy}) is of
order $\mathcal{O}(t^0)$ and takes the form
\begin{eqnarray}
\bigtriangleup \varphi _0(\vec{x}) +C\varphi _0(\vec{x})=0\,,
\end{eqnarray}
where $C$ is a constant given by $-6A_1^2-12A_2$. Thus, only one
specific Fourier mode of $\varphi _0$ could be consistently transferred to
the visible brane. This certainly is inconsistent with obtaining
a scale invariant spectrum on the brane. Thus, we will not further
consider this possibility. Therefore, we conclude that only the trivial
solution
\begin{eqnarray}
\varphi _{0}(\vec{x})&\equiv&0\,,\label{result1}\\
\psi _{0}(\vec{x})&\equiv&0 \label{result2}
\end{eqnarray}
is consistent with our model if $k\geq 3$. This is our first main result.
Its implications will be discussed
in section \ref{sectionconsequences}.

\subsubsection{The Einstein Equations for $k=1$ \label{sectionk=1}}
Let us consider the case $k=1$, that is the metric
(\ref{metrikafterbounce1})-(\ref{metrikafterbounce3}).
The $x_i$-$t$ equation (\ref{einsteinxit}), evaluated to order
$\mathcal{O}(t^{-1})$, yields
\begin{eqnarray}
\psi _0(\vec{x})=-2\varphi _0(\vec{x})\,.
\end{eqnarray}
Since the junction conditions (\ref{constraint1}) and (\ref{constraint2})
imply $\psi _0(\vec{x})=-\varphi _0(\vec{x})$
if $s>1$ and/or $q>1$, we must have $s=q=1$ and the boundary conditions
\begin{eqnarray}
\varphi _1^{\prime}(\vec{x},y=y_i)&=&a _1^{\prime}(y=y_i)
\varphi _{0}(\vec{x})\,,\label{bc1}\\
\psi _1^{\prime}(\vec{x},y=y_i)&=&a _1^{\prime}(y=y_i)
\varphi _{0}(\vec{x})\,,\label{bc2}\\
\xi _1^{\prime}(\vec{x},y=y_i)&=&6a _1^{\prime}(y=y_i)
\varphi _{0}(\vec{x})\,.\label{bc3}
\end{eqnarray}
The $y$-$x_i$ equation (\ref{einsteinyxi}) to order
$\mathcal{O}(t^{-1})$ yields
\begin{eqnarray}
2\psi _1^{\prime}+\varphi _1^{\prime}-3a_1^{\prime}\varphi _0
+\frac{3}{2}w_2=0\label{e1eom1}\,.
\end{eqnarray}
Evaluating this at the brane gives us the boundary condition
\begin{eqnarray}
w_2 (\vec{x},y=y_i)=0\,.\label{bc4}
\end{eqnarray}
Another equation of motion follows from the $t$-$t$ equation
(\ref{einsteintt}) to order $\mathcal{O}(t^{-1})$
\begin{eqnarray}
\frac{\psi _1^{\prime\prime}}{B^2}=\psi _1 \label{e1eom2}\,,
\end{eqnarray}
which is solved by
\begin{eqnarray}
\psi _1(\vec{x},y)=a_1(y)\varphi _0(\vec{x})\,,
\end{eqnarray}
where we made use of (\ref{bc2}). The $x_i$-$t$ equation
(\ref{einsteinxit}) to order $\mathcal{O}(t^0)$ gives
yet another equation of motion
\begin{eqnarray}
\varphi _1+a_1\varphi _0+\frac{1}{6B^2}w_2^{\prime}=0\,,
\end{eqnarray}
which can be combined with (\ref{e1eom1}) to yield
\begin{eqnarray}
\frac{\varphi _1^{\prime\prime}}{B^2}=9\varphi _1+10a_1\varphi _0\,,\\
\frac{w_2^{\prime\prime}}{B^2}=9w_2-12a_1^{\prime}\varphi _0\,.
\end{eqnarray}
These are solved by
\begin{eqnarray}
\varphi _1(\vec{x},y)&=&-\frac{5}{4}a_1(y)\varphi _0(\vec{x})
+\frac{3}{4}\tilde{A}_1\varphi _0(\vec{x})\\
\nonumber &&\!\times\!
\left(\frac{1-\cosh(3By_2)}{\sinh(3By_2)}\cosh(3By)+\sinh(3By)\right),\\
w_2(\vec{x},y)&=&\frac{3}{2}B\tilde{A}_1\Bigg[
\cosh(3By)-\cosh(By)\\
\nonumber &&+\frac{1-\cosh(3By_2)}{\sinh(3By_2)}\sinh(3By)
\\
\nonumber &&-\frac{1-\cosh(y_2B)}{\sinh(y_2B)}\sinh(yB)\Bigg]\varphi _0(\vec{x}),
\end{eqnarray}
where we made use of the boundary conditions (\ref{bc1}) and (\ref{bc4}).
For $\xi _1$ we get from (\ref{eomphi1}), to leading order in $t$,
\begin{eqnarray}
\frac{\xi _1^{\prime\prime}}{B^2}=\xi _1+12a_1\varphi _0\,,
\end{eqnarray}
with the solution
\begin{eqnarray}
\xi _1(\vec{x},y)&=&\Bigg[3(\tilde{A}_1+A_1)
\left(By+C_1-\frac{3}{2}\right)e^{By}\\
\nonumber &&\,\,\,3(\tilde{A}_1+A_1)
\left(By+C_2+\frac{3}{2}\right)e^{-By}\Bigg]\varphi _0(\vec{x})\,,
\end{eqnarray}
where $C_1$ and $C_2$ are constants chosen such that (\ref{bc3}) is
satisfied. A nontrivial consistency check for these
solutions is, e.g., the $x_i$-$x_i$ equation (\ref{einsteinxixi})
evaluated to order $\mathcal{O}(t^{-1})$, which is indeed satisfied.

Now we have to look at all other Einstein equations up to order
$\mathcal{O}(t^0)$, all of which
include higher order perturbations. In order for a scale-invariant
spectrum to be consistent,
the terms $\sim\bigtriangleup \varphi _0$ in (\ref{einsteinyy}) and
(\ref{einsteintt}) should get canceled
for all $y$. We can achieve this if we write
\begin{eqnarray}
\psi _2(\vec{x},y)\equiv\varphi _0(\vec{x})\psi _2(y)+\tilde{\psi}_2(\vec{x})
\end{eqnarray}
and demand that
\begin{eqnarray}
\tilde{\psi}_2(\vec{x})=\frac{1}{2}\bigtriangleup \varphi _0(\vec{x})\,.
\end{eqnarray}
Combining (\ref{einsteinyt}) to order $\mathcal{O}(t^{-1})$ with
(\ref{einsteintt}) and (\ref{einsteinyy}) to order
$\mathcal{O}(t^0)$ gives three equations of motion for $a_2(y),n_1(y)$
and $\psi _2(y)$. Together with (\ref{eomb2})
these could be solved. Since $a_2$ is a solution to the background equations
involving the brane interaction,
one could in principle extract information about the brane interaction by
demanding consistency in the
perturbed metric beyond leading order. Since we did not specify the brane
interaction so far, we shall be satisfied
with consistency to leading order.

The remaining three Einstein equations (\ref{einsteinyt}),
(\ref{einsteinyxi}) and (\ref{einsteinxixi}), all
evaluated to order $\mathcal{O}(t^0)$, involve six new degrees of freedom:
$\psi _3,\varphi _2,w_3,n_2,b_3$ and $a_3$.
These define $\psi _3,\varphi _2$ and $w_3$ in terms of higher order terms
of the metric, all of which require
a better understanding of the brane interaction.

All that is left to do now is to check if we can impose the initial conditions
\begin{eqnarray}
\varphi _0(\vec{x})=-\varphi _1(\vec{x},y=0)t_i\,,\\
\psi _0(\vec{x})=-\psi _1(\vec{x},y=0)t_i\,,
\end{eqnarray}
such that at $t=t_i$ perturbations $\Phi$ and $\Psi$ on the visible brane
vanish up to order $\mathcal{O}(t_i^2)$ (they do not vanish
on the hidden brane).
We can arrange this if we assume $By_2>>1$,
$-\alpha _1e^{-\Xi _0}\geq 8$ -- then $t_i\geq -1/2$
and
\begin{eqnarray}
\Phi(\vec{x},y=y_2,t=t_i)=\frac{3}{2}\varphi _0(\vec{x})+\mathcal{O}(t_i^2)\,,\\
\Psi(\vec{x},y=y_2,t=t_i)=-4\varphi _0(\vec{x})+\mathcal{O}(t_i^2)\,.
\end{eqnarray}
It should be noted that in order for this to be possible we need the
visible brane to be a negative tension brane.
Since at $t=0$ the perturbations on the visible brane are given
by $\Phi=\varphi _0$ and $\Psi=-2\varphi _0$, the above two
equations tell us how to match perturbations shortly before the bounce
in the bulk (and on the hidden brane) to perturbations
on the visible brane at the bounce.

The only change to our setup after the bounce is a change in the junction
conditions in section \ref{sectionconstraints}
due to the matter which has been generated by the bounce on the visible
brane. Therefore, a consistent solution after the bounce could be found
along similar lines as above.

\section{Consequences for the Ekpyrotic and Cyclic Scenarios
\label{sectionconsequences}}

We have seen that for $k\geq 3$ the only solution to the
perturbation equations that is bounded near the bounce
and does not yield a singularity in the Riemann tensor is the
trivial solution (\ref{result1}) and (\ref{result2}) --
the solution that has vanishing metric perturbations on the
branes at the bounce.

Since there have to be some perturbations on the visible brane after the
bounce if the Ekpyrotic Universe is to describe our Universe,
we are led to conclude at least one of the following
\begin{enumerate}
\item The background Riemann tensor and/or energy-momentum tensor
does not remain bounded at the bounce and thus we have a singularity
even from the 5D point of view.
\item At least one perturbation variable exhibits a growing mode
$\sim t^{-c}$ where $c>0$, and thus perturbation theory breaks down as we
approach the bounce.
\end{enumerate}
In either case the bounce can not be examined using standard methods.
One has to specify the type of the singularity and find a way to match
perturbations
shortly before and shortly after the bounce. Unfortunately the
breakdown of first order perturbation theory
renders the whole problem highly gauge dependent.
The most recent attempt addressing this problem \cite{Tolley2} assumes a
Milne-like Universe and works in a specific gauge, but the robustness of the
results towards changes in the specific assumptions concerning
the background and the matching conditions is subject of an ongoing debate.

However, in the case that the two boundary branes approach each other with
constant velocity near the bounce, that is $k=1$, a consistent
solution of the perturbed Einstein equations can be found, as was done
explicitly in section \ref{sectionk=1}. In this
case, one can easily follow the perturbations until the actual collision.
Furthermore, if a scale invariant
spectrum in the bulk was generated while the hidden brane was approaching the
visible brane, then it will indeed be transferred
to the visible brane during the collision. Since the bounce is non-singular and
all perturbations remain small, one needs neither to worry about higher
curvature corrections nor gauge ambiguities.
In order to see if this single case is actually realized one needs a
better understanding of the brane interaction and
check consistency beyond leading order in the perturbed metric.

The limitations of our conclusions lie in the specific realization of the
model used. Our choice was guided by
the principle of simplicity, apparent e.g. in the choice of the bulk fields.
Furthermore we did not
specify the brane interaction -- we only assumed that it vanishes at least
like $t$ during the bounce.
We showed that in this simplified setup
a consistent realization of the Ekpyrotic or Cyclic scenarios
including perturbations is possible.

\begin{acknowledgments}

We wish to thank Syksy R\"as\"anen for useful correspondence,
and Paul Steinhardt for comments on the first version of this
paper. One of us
(SP) wishes to acknowledge Jean-Luc Lehners at Imperial College for
discussion concerning brane perturbations, and the Imperial College
Theory Group for the use of their facilities in the course of the
work on this paper. RB wishes to acknowledge the hospitality of the
McGill Physics Department. This work was supported in part by the
U.S. Department of Energy under Contract DE-FG02-91ER40688, TASK A.

\end{acknowledgments}

\end{document}